\documentstyle[preprint,prc,aps,epsfig]{revtex}
\tightenlines
\begin{document}
\title{A Global Potential Analysis of the $^{16}$O+$^{28}$Si
Reaction Using a New Type of Coupling Potential}
\author{I. Boztosun \footnote[2]{Present
address: Computational Mathematics Group, School of Computer
Science and Mathematics, University of Portsmouth, Portsmouth PO1
2EG UK}}
\address{Department of Physics, Erciyes University, 38039 Kayseri Turkey}
\author{W.D.M. Rae}
\address{Department of Nuclear Physics, University of Oxford,
Keble Road, Oxford OX1 3RH UK}
\date{\today}
\maketitle
\begin{abstract}
A new approach has been used to explain the experimental data for
the $^{16}$O+$^{28}$Si system over a wide energy range in the
laboratory system from 29.0 to 142.5 MeV. A number of serious
problems has continued to plague the study of this system for a
couple of decades. The explanation of anomalous large angle
scattering data; the reproduction of the oscillatory structure
near the Coulomb barrier; the out-of-phase problem between
theoretical predictions and experimental data; the consistent
description of angular distributions together with excitation
functions data are just some of these problems. These are long
standing problems that have persisted over the years and do
represent a challenge calling for a consistent framework to
resolve these difficulties within a unified approach. Traditional
frameworks have failed to describe these phenomena within a single
model and have so far only offered different approaches where
these difficulties are investigated separately from one another.
The present work offers a plausible framework where all these
difficulties are investigated and answered. Not only it improves
the simultaneous fits to the data of these diverse observables,
achieving this within a unified approach over a wide energy range,
but it departs for its coupling potential from the standard
formulation. This new feature is shown to improve consistently the
agreement with the experimental data and has made major
improvement on all the previous coupled-channels calculations for
this system.
\end{abstract}
%
%
\begin{center}
 \end{center}
\newpage
\section{Introduction}
Since the first observation of the unexpectedly large
cross-section near $\theta_{CM}$=180$^{\circ}$ for the elastic
and the inelastic scattering between light and medium heavy nuclei
\cite{Bra77}, considerable experimental and theoretical efforts have
been devoted to the systematic studies of this phenomenon and related aspects.

The physical origin of the observed structure is not yet fully
understood \cite{Bra82,Boz99} and presents a challenge to
different approaches that have been proposed to explain it. These
approaches range from the occurrence of possibly overlapping shape
resonances \cite{Bar78} and the scattering from
surface-transparent optical potentials \cite{Kah79} to more exotic
effects like explicit parity dependence of the ion-ion potential
\cite{Deh78,Kub79}. At present, none of these approaches provides
a consistent explanation for all the existing data for this
system.

Consequently, the following problems continue to exist for this
reaction \cite{Boz1,Boz2,Bozth,Bozlett}: $(1)$ the explanation of
anomalous large angle scattering data; $(2)$ the reproduction of
the oscillatory structure near the Coulomb barrier; $(3)$ the
out-of-phase problem between theoretical predictions and
experimental data; $(4)$ the consistent description of angular
distributions together with excitation functions data; $(5)$ the
deformation parameters ($\beta$ values): previous calculations
require $\beta$ values that are at variance with the empirical
values and are physically unjustifiable.

The elastic and inelastic scattering data of the
$^{16}$O+$^{28}$Si system have been studied extensively and some
of the above-mentioned problems could not be accounted for
\cite{Kob84,Bra85,Sci97,Fil89,Sto79,Bra97}. The most extensive
study for this system was carried out by Kobos and Satchler
\cite{Kob84} who used a double folding potential with two small
additional {\it ad-hoc} potentials to reproduce the measured
elastic scattering data. Without two small additional potentials,
they observed that the theoretical calculations and the
experimental data were completely out-of-phase and could not
reproduce the experimental data.

Therefore, building on two previous papers \cite{Boz1,Boz2}, which
were outstandingly successful in explaining the experimental data
for the $^{12}$C+$^{12}$C and $^{12}$C+$^{24}$Mg reactions which
both have been intensively investigated over the
years~\cite{Sci97,Fil89,Sto79,Bra97,Car76,Car78}, we investigate
the elastic and inelastic scattering of $^{16}$O+$^{28}$Si system
from 29.0 MeV to 142.5 MeV. The excitation functions for the
ground and the first excited states have also been analyzed over
this energy range. In this paper, our aim is to reproduce all the
experimental data with empirical $\beta$ value.

In the next section, we first introduce the standard
coupled-channels model and then show the results of these analyzes
in section \ref{stanres} from E$_{Lab}$=29.0 MeV to 142.5 MeV. In
section \ref{newcc}, we introduce a new coupling potential to
analyze the experimental data in the same energy range and show
the results of these new coupled-channels calculations. Finally,
section \ref{conc} is devoted to our summary and conclusion.
\section{The Standard Coupled-Channels Calculations}
\label{stan}
In the present coupled-channels calculations, we
describe the interaction between $^{16}$O and $^{28}$Si nuclei
with a deformed optical potential. The real potential is assumed
to have the square of a Woods-Saxon shape:
\begin{equation}
V_{N}(r) = \frac{-V_{0}}{(1+exp(r-R)/a)^{2}} \label{realpot}
\end{equation}
with $V_{0}$=706.5 MeV, R=$r_{0}$($A_{P}$$^{1/3}$+$A_{T}$$^{1/3}$)
with $r_{0}$=0.7490 fm  and a=1.40 fm. The parameters of the real
potential are fixed as a function of energy and are not changed in
the present calculations although it was observed that small
changes could improve the quality of the fits. The Coulomb
potential with a radius of 5.56 fm is also added.

\begin{figure}[bt]
\epsfxsize 10.5cm \centerline{\epsfbox{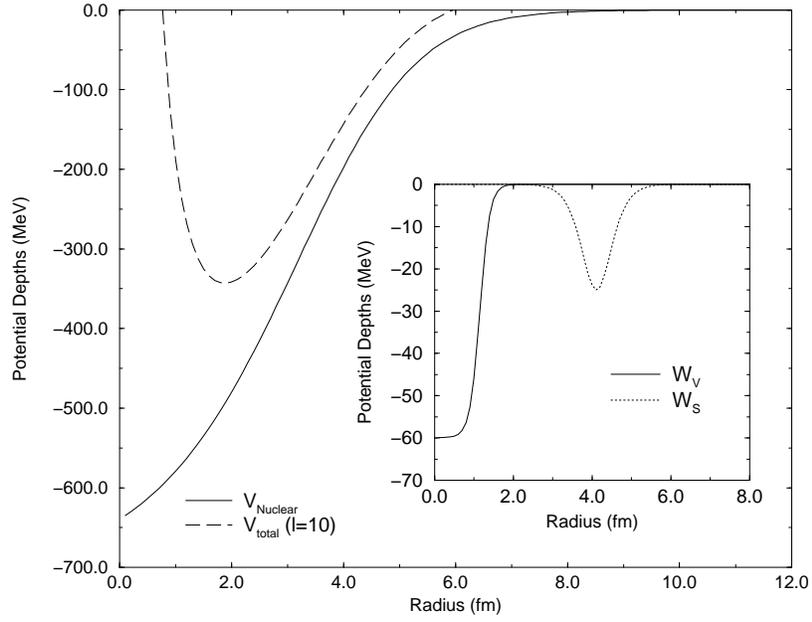}} \vskip+1.0cm
\caption{The real and imaginary parts of the potential between
$^{16}$O and $^{28}$Si are plotted against the separation R for
$l$=10. $W_{V}$ denotes the volume and $W_{S}$ the surface
components of the imaginary potential at $E_{Lab}$=41.17 MeV (the
inserted figure).} \label{pot}
\end{figure}

The imaginary part of the potential is taken as the sum of a
Woods-Saxon volume and surface potentials:
\begin{equation}
W(r)=-W_{V}f(r,R_{V},a_{V})+4W_{S}a_{S}df(r,R_{S},a_{S})/dr
 \label{imagpot}
\end{equation}
\begin{equation}
f(r,R,a) = \frac{1}{1+exp((r-R)/a)}
\end{equation}
with $W_{V}$=59.9 MeV, $a_{V}$=0.127 fm and $W_{S}$=25.0 MeV,
$a_{S}$=0.257 fm. These parameters are also fixed in the
calculations and only their radii increase linearly with energy
according to the following formulae:
\begin{equation}
R_{V} = 0.061E_{CM}-0.44 \label{volr}
\end{equation}
\begin{equation}
R_{S} = 0.241E_{CM}-2.19 \label{surr}
\end{equation}
The real and imaginary potentials are shown in figure \ref{pot}
for $E_{Lab}$= 41.17 MeV. The sum of the nuclear, Coulomb and the
centrifugal potentials is also shown in the same figure for the
orbital angular momentum quantum number, $l=10$. The superposition
of the attractive and repulsive potentials results in the
formation of a potential pocket, which the width and depth of the
pocket depend on the orbital angular momentum. This pocket is very
important for the interference of the barrier and internal waves,
which produces the pronounced structure in the cross-section. The
effect of this pocket can be understood in terms of the
interference between the internal and barrier waves that
correspond to a decomposition of the scattering amplitude into two
components, the inner and external waves~\cite{Lee78,Bri77}.

The relative significance of the volume and surface components of
the imaginary potential has also been examined for all the
energies considered. For higher energies, omitting the volume term
predominantly affects the amplitude of the cross-section at large
angles. However, this effect is small and negligible at lower
energies. Omitting the surface term increases the cross-sections
at large angles which are as much as two orders of magnitude. It
is observed that this term has a significant effect at all the
energies considered.

Since the target nucleus $^{28}$Si is strongly
deformed, it is essential to treat its collective
excitation explicitly in the framework of the coupled-channels
formalism. It has been assumed that the target nucleus has a static
quadrupole deformation, and that its rotation can be described in
the framework of the collective rotational model. It is therefore
taken into account by deforming the real optical potential in the following
way
\begin{equation}
R(\theta,\phi)=r_{0}A_{P}^{1/3}+r_{0}A_{T}^{1/3}[1+\beta_{2}
Y_{20}(\theta,\phi)]
\end{equation}
where $P$ and $T$ denote the projectile and target nuclei
respectively and $\beta_{2}$=-0.64 is the deformation parameter of
$^{28}$Si. This value is actually larger than the value calculated
from the known B(E2). However this larger $\beta_{2}$ was needed
to fit the magnitude for the 2$^{+}$ state data as discussed in
the next sections.

In the present calculations, the first two excited states of the
target nucleus $^{28}$Si, {\it i.e.} 2$^{+}$ (1.78 MeV) and
4$^{+}$ (4.62 MeV), are included and the 0$^{+}$-2$^{+}$-4$^{+}$
coupling scheme is employed. The reorientation effects for 2$^{+}$
and 4$^{+}$ excited states are also included. The inclusion of the
2$^{+}$ and 4$^{+}$  excited states has important effects as their
effects change the elastic scattering fits substantially. These
effects confirm that it is essential to use the coupled-channels
method in the case where one of the nuclei in the reaction is
strongly deformed. Extensively modified version of the code {\it
CHUCK} \cite{Kunz} has been used for the all calculations.
\section{Results}
\label{stanres}
\begin{figure}[bt]
\epsfxsize 13.5cm \centerline{\epsfbox{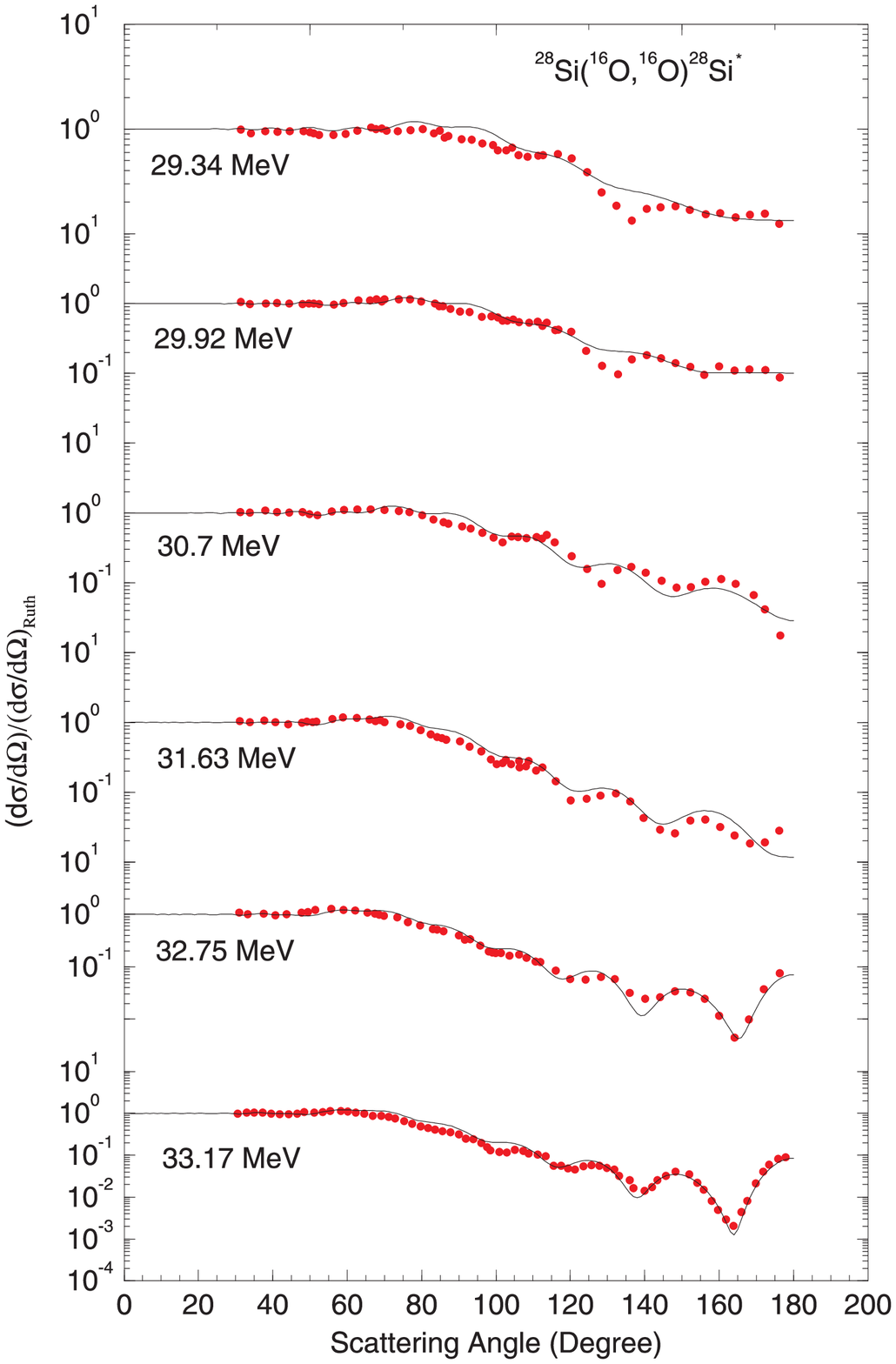}} \vskip+0.5cm
\caption{Ground state results obtained using the standard
coupled-channels model with $\beta_{2}$=-0.64.}
\label{ws2ground1osi}
\end{figure}

\begin{figure}[bt]
\epsfxsize 14.5cm \centerline{\epsfbox{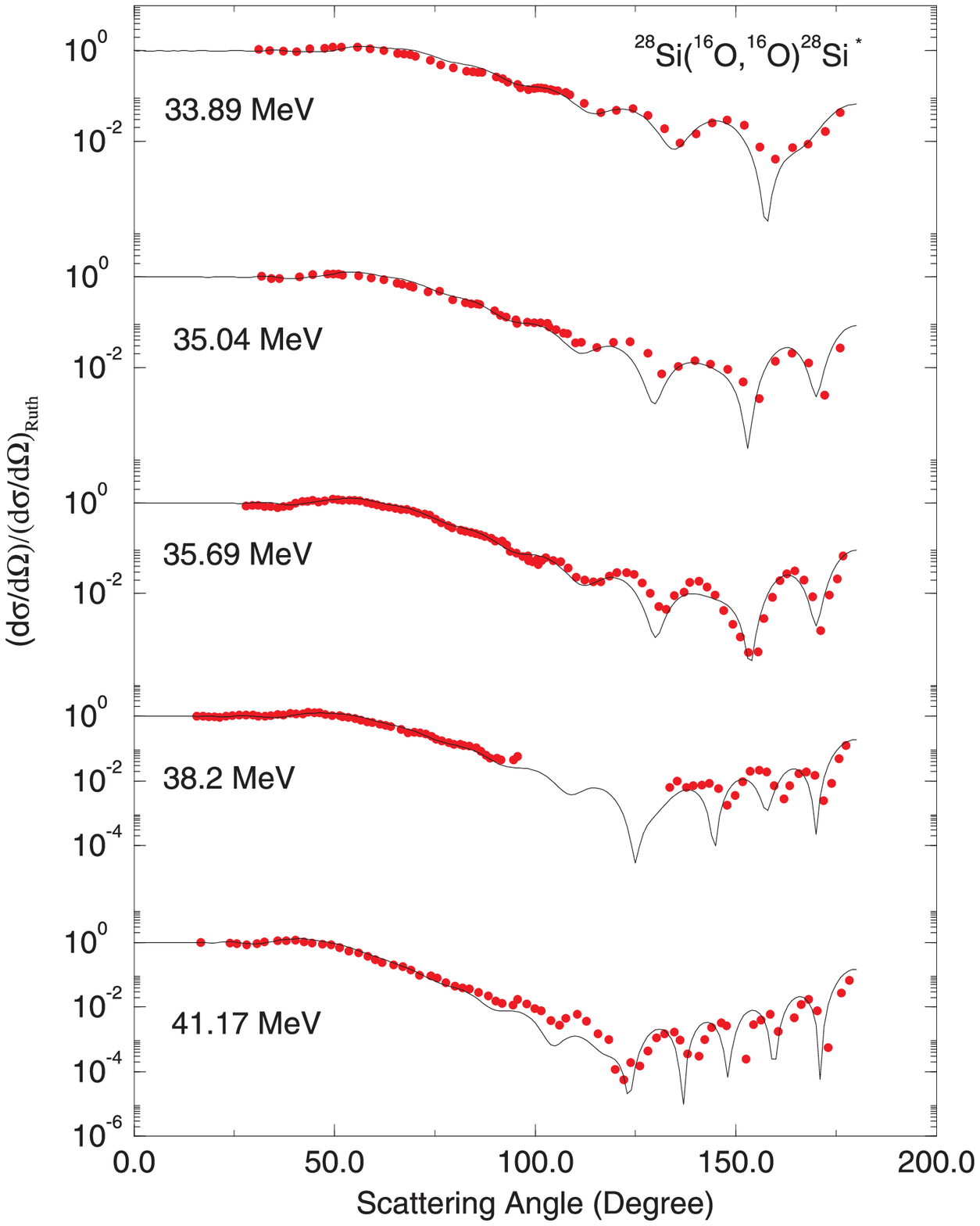}} \vskip+0.5cm
\caption{Ground state results obtained using the standard
coupled-channels model with $\beta_{2}$=-0.64 ({\it continued from
figure \ref{ws2ground1osi}}).} \label{ws2ground2osi}
\end{figure}

\begin{figure}[bt]
\epsfxsize 9.5cm \centerline{\epsfbox{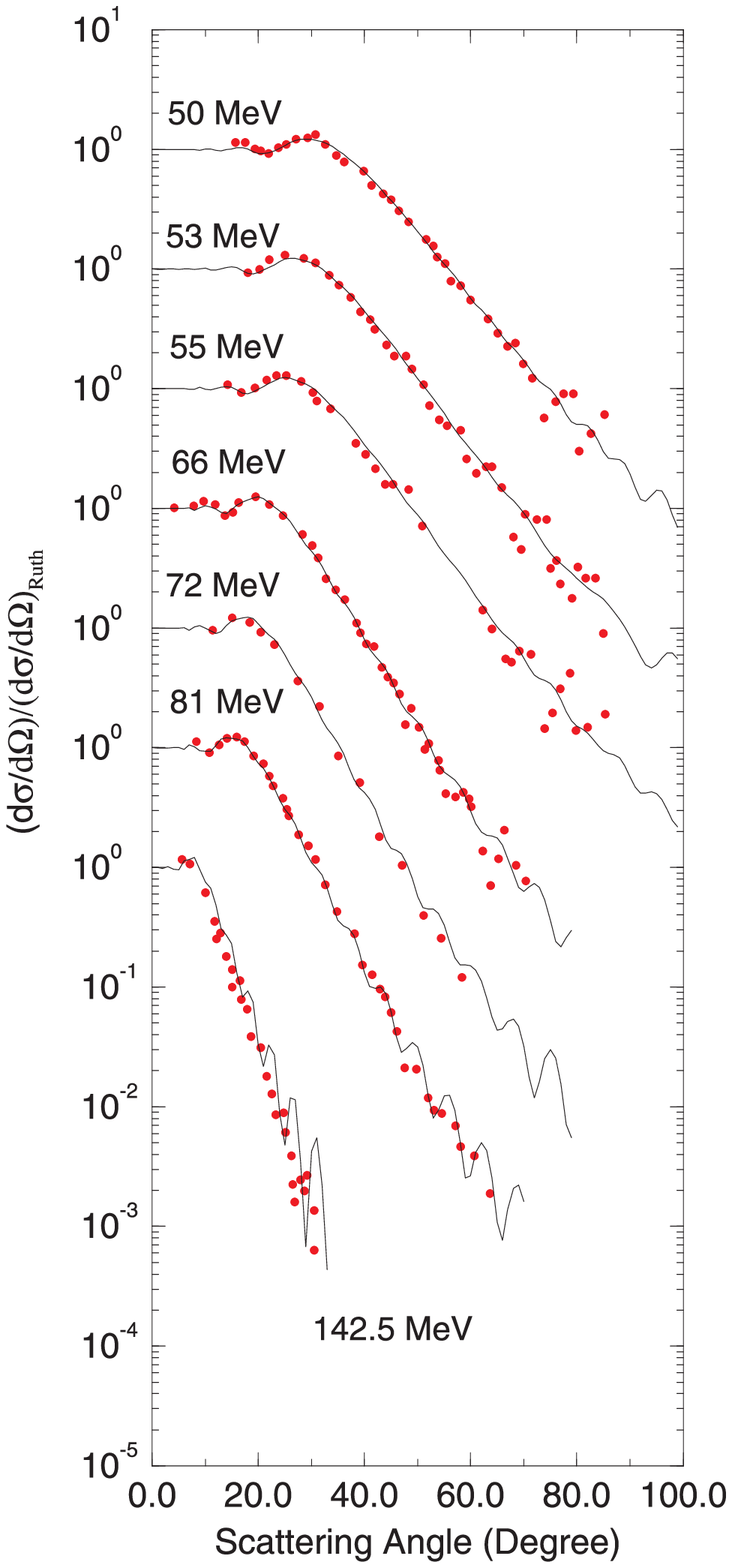}} \vskip+0.5cm
\caption{Ground state results obtained using the standard
coupled-channels model with $\beta_{2}$=-0.64 ({\it continued from
figure \ref{ws2ground2osi}}).} \label{ws2ground3osi}
\end{figure}

\begin{figure}[bt]
\epsfxsize 13.5cm \centerline{\epsfbox{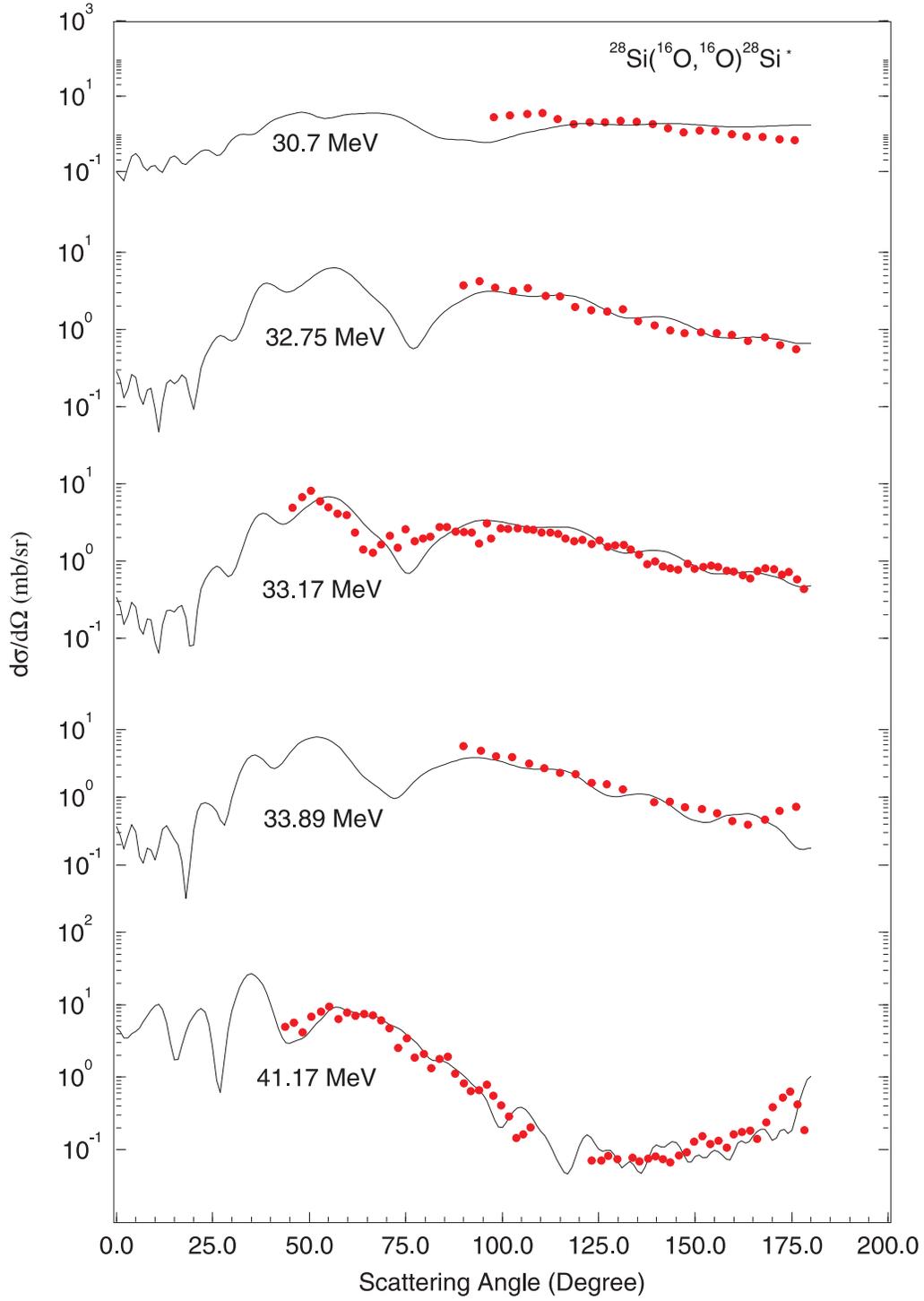}} \vskip+0.5cm
\caption{2$^{+}$ excited state results obtained using the standard
coupled-channels model with $\beta_{2}$=-0.64.} \label{ws2in}
\end{figure}

Using this standard coupled-channels model, the results for the
ground and first excited states are shown in figures
\ref{ws2ground1osi}, \ref{ws2ground2osi}, \ref{ws2ground3osi} and
\ref{ws2in}. It should be stressed that very close fits to the
experimental data at forward, backward and intermediate angles
were obtained without applying any {\it ad-hoc} procedures other
than increasing the $\beta_{2}$ value (see $\chi^{2}$ values in
table~\ref{chi}). In general, the previous coupled-channels
calculations aiming to explain the structures at large angles
obtained rather poor fits at forward angles or vice versa. Even
when the forward and backward angles were fitted, the intermediate
angles were not \cite{Bra77,Dud78}.

However, there are problems in our first excited state results.
The magnitude of the cross-sections and the phase of the
oscillations are obtained correctly at most angles. However, when
one looks at the 2$^{+}$ state results in detail, it is apparent
that the experimental data and our predictions are out-of-phase
towards large angles at higher energies. This problem was also
found in earlier coupled-channels calculations for this
system~\cite{Bra77,Bra85,Dud78}.

When studying this reaction systematically in a wide energy range,
we came across several problems: The first problem relates to the
oscillatory structure and to the backward rise in the
cross-section at large angles for which the standard
coupled-channels model  provides a solution.

The second problem pertains to the calculation of the first excited
state cross-section. Using the exact $\beta$ value, we observed that
the calculations underestimated the experimental data, a phenomenon
confirmed by other works which assert that in
the coupled-channels or DWBA calculations, one has to increase or
decrease the deformation parameter($\beta$) to be able to get agreement
with the measured experimental data and that the choice of the $\beta$ value
is somehow arbitrary in fitting the data. Therefore,
we also adopted to increase the $\beta$ value.

The third problem arises due to Blair's phase rule \cite{Sat83}
which states that `the oscillations for even-$l$ transfer are
out-of-phase with those for elastic scattering, while those for
odd-$l$ transfer are in phase'. These experimental data obey this
rule at numerous energies, except the energies around
$E_{Lab}$$\sim$35.0 MeV (see figures \ref{ws2ground2osi} and
\ref{ws2in}). While the measured cross-section for the ground
state has maxima at $\sim$180$^{\circ}$, it has also maxima at
$\sim$180$^{\circ}$ for the 2$^{+}$ state whereas, it should have
minima $\sim$180$^{\circ}$ according to the Blair's phase rule.

\begin{figure}[bt]
\epsfxsize 9.5cm \centerline{\epsfbox{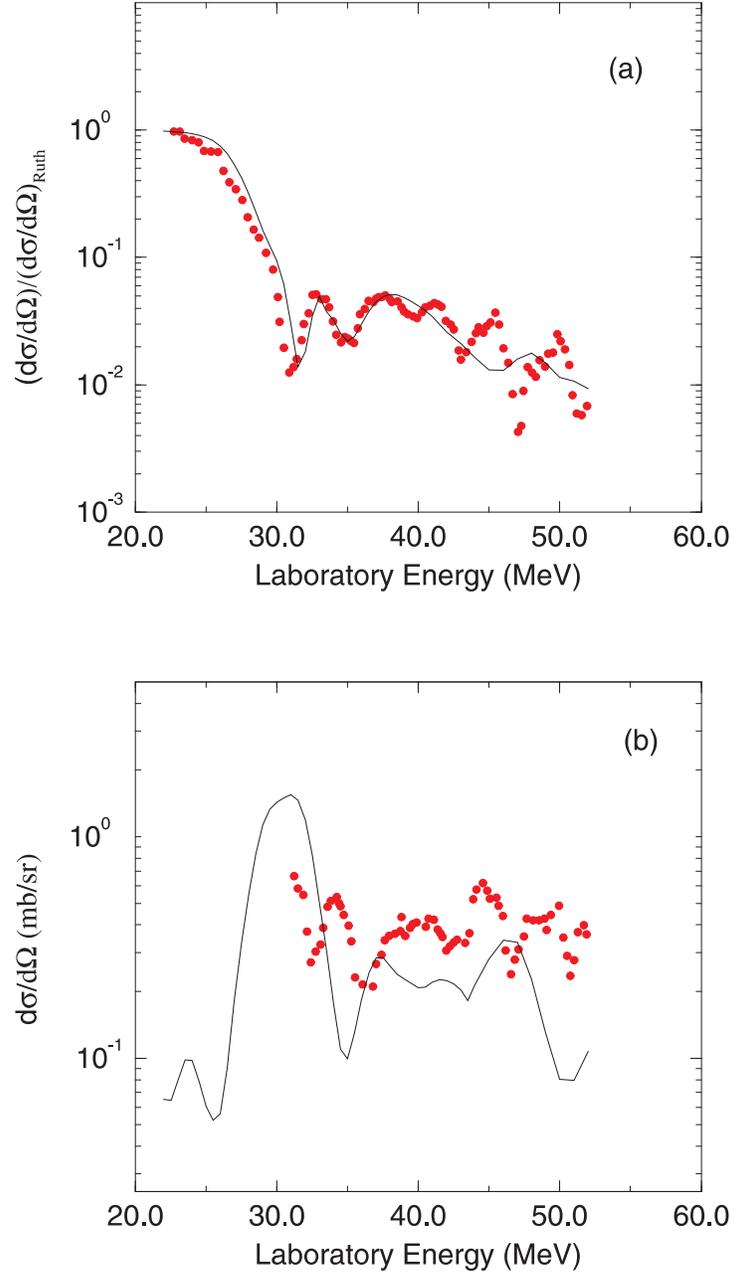}} \vskip+0.5cm
\caption{180$^{\circ}$ excitation function results obtained using
the standard coupled-channels model for $(a)$ the ground and (b)
2$^{+}$ states with $\beta_{2}$=-0.64.} \label{ws2exc}
\end{figure}

The theoretical predictions are completely out-of-phase around
these energies for the 2$^{+}$ state although they fit the ground
state data. This problem is also clearly observed in the
180$^{\circ}$ excitation function of the 2$^{+}$ state as shown in
figure \ref{ws2exc}. The magnitudes of our calculations are also
at least twice bigger than the experimental data at lower
energies.

Our coupled-channels calculations showed that the threshold energy
is $E_{Lab}$$\sim$35.0 MeV. It was also reported \cite{Mar 99}
that the irregular behavior of the experimental data starts beyond
this energy. No coupled-channels calculation has been carried out
below and above this energy simultaneously. However, research has
been conducted and studies have been published pertaining to below
or above this energy ($E_{Lab}$$\sim$35.0 MeV).

In the past, a number of models have been proposed in order to
solve the above-mentioned problems, ranging from isolated
resonances \cite{Bar78,San80} to cluster exchange between the
projectile and target nucleus \cite{Fil89,Fra79} (see \cite{Bra82}
for a detailed discussion). We have attempted to overcome these
problems by modifying the shape and the parameters of the real
potential as well as the parameters and the shape of the imaginary
potential. These modifications in the real and imaginary
potentials improved the 180$^{\circ}$ excitation function.
However, we were unable to fit individual angular distributions
and excitation functions simultaneously over the whole energy
range.

We then sought to include the 6$^{+}$ excited state. The inclusion
of one additional excited state weakened the imaginary potential
and this was useful to infer what the shape of the imaginary
potential should be. Nevertheless, we were unable to include it in
the final stage since it created a numerical accuracy/instability
problem in the code. We finally changed the $\beta_{2}$ value and
included a $\beta_{4}$ deformation. However, varying the value of
the $\beta_{2}$ and the inclusion of the $\beta_{4}$ did not solve
the problems.

In summary, these attempts failed to provide a wholistic
solution to the above-mentioned problems.
We were unable to explain the elastic and inelastic scattering
data as well as their 180$^{\circ}$ excitation functions simultaneously.
\section{New Coupling Potential}
\label{newcc}
The limitations of the standard coupled-channels theory in the
analysis of this reaction has been well established by both our
analyzes and the works published so far. We came across the same
type of failure of the standard coupled-channels method in
explaining the experimental data for the $^{12}$C+$^{12}$C and
$^{12}$C+$^{24}$Mg reactions.

In order to explain the experimental data for these systems, we
had to introduce a new type of coupling potential, which is a
second-derivative coupling potential used in the place of the
usual first derivative coupling potential. This new coupling
potential has successfully explained the scattering observables of
these two reactions over wide energy ranges and has made major
improvement on the all the previous coupled-channels calculations
for these systems. The reason and a possible interpretation of
such a new coupling potential have been discussed in the two
previous papers \cite{Boz1,Boz2}.

\begin{figure}[bt]
\epsfxsize 10.5cm \centerline{\epsfbox{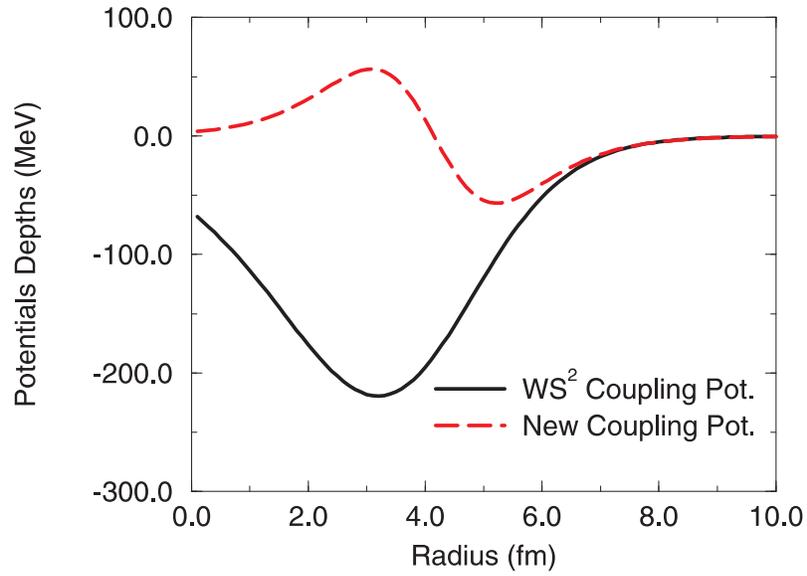}} \vskip+0.5cm
\caption{The comparison of the {\it standard} coupling potential
and our {\it new} coupling potential, parameterized as the
2$^{nd}$ derivative of Woods-Saxon shape.} \label{coupling}
\end{figure}

Building on these two previous papers, here we use a new
second-derivative coupling potential to find a global solution to
the problems that $^{16}$O+$^{28}$Si reaction manifests. This new
coupling potential is displayed in comparison with the standard
coupling potential in figure \ref{coupling}. It is parameterized
accurately as the second-derivative of the Woods-Saxon shape in
the following form:

\begin{equation}
V_{C}(r) = \frac{-V_{C_{0}} \,\, e^{x}
(e^{x}-1)}{a^{2}\left[1+e^{x} \right]^{3}} \label{couppot}
\end{equation}
where $x=(r-R)/a$ and $V_{C_{0}}$=155.0 MeV, $R$=4.16 fm and
$a$=0.81 fm.

In the new coupled-channels calculations, the real and imaginary
potentials have the same shapes as given by the equations
\ref{realpot} and \ref{imagpot} and the same potential parameters
are used except for the depth of the real potential and the
$\beta_{2}$ value. They have to be readjusted as $V_{0}$=750.5 MeV
and $\beta_{2}$=-0.34, which corresponds to the exact value
derived from the life time of the 2$^{+}$ state
\cite{Dud78,End73,Sch77}.

\begin{figure}[bt]
\epsfxsize 13.5cm \centerline{\epsfbox{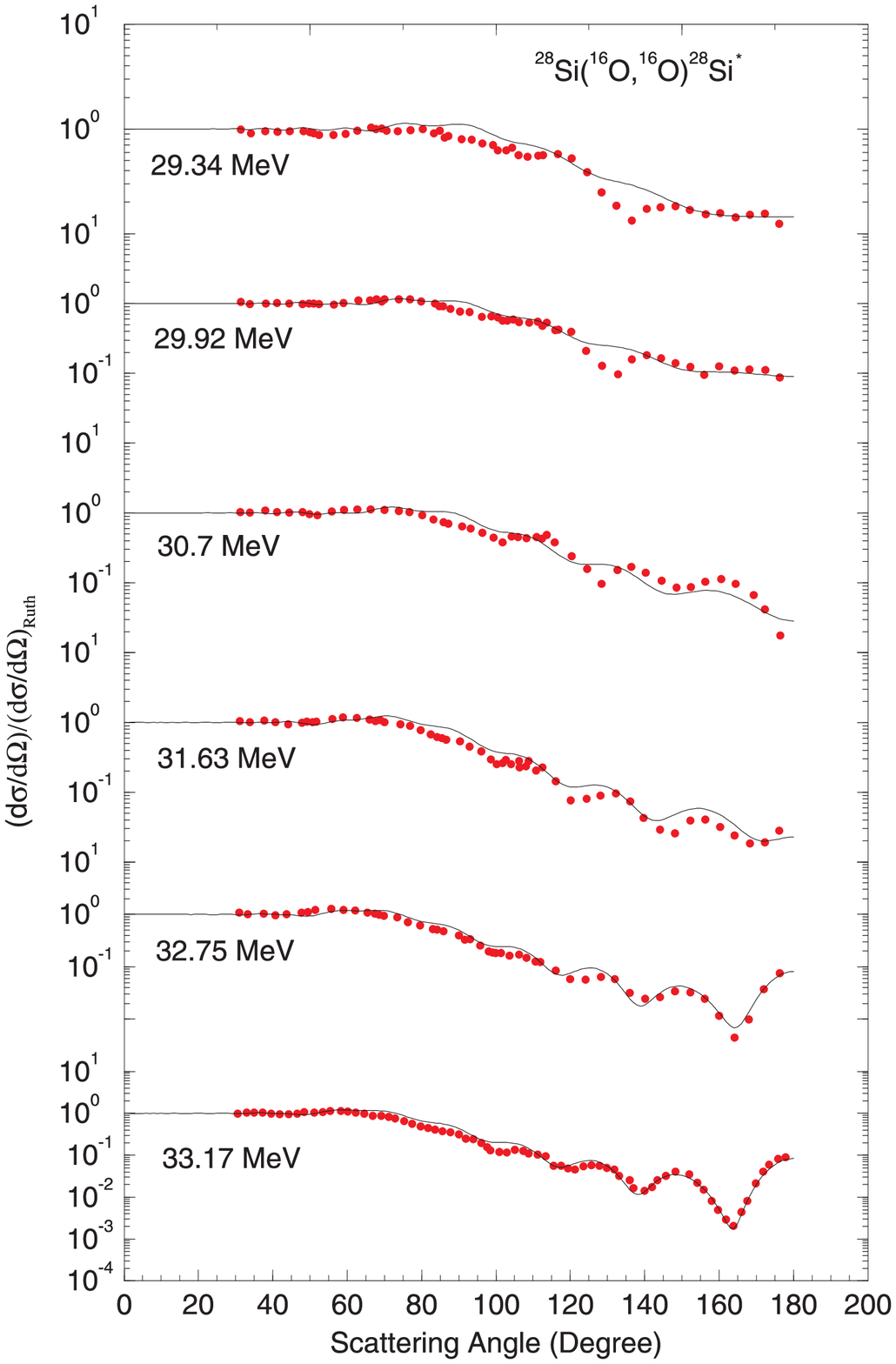}} \vskip+0.5cm
\caption{Ground state results obtained using the new coupling
potential with the exact $\beta$ value.} \label{2ndgroundosi}
\end{figure}

\begin{figure}[bt]
\epsfxsize 14.5cm \centerline{\epsfbox{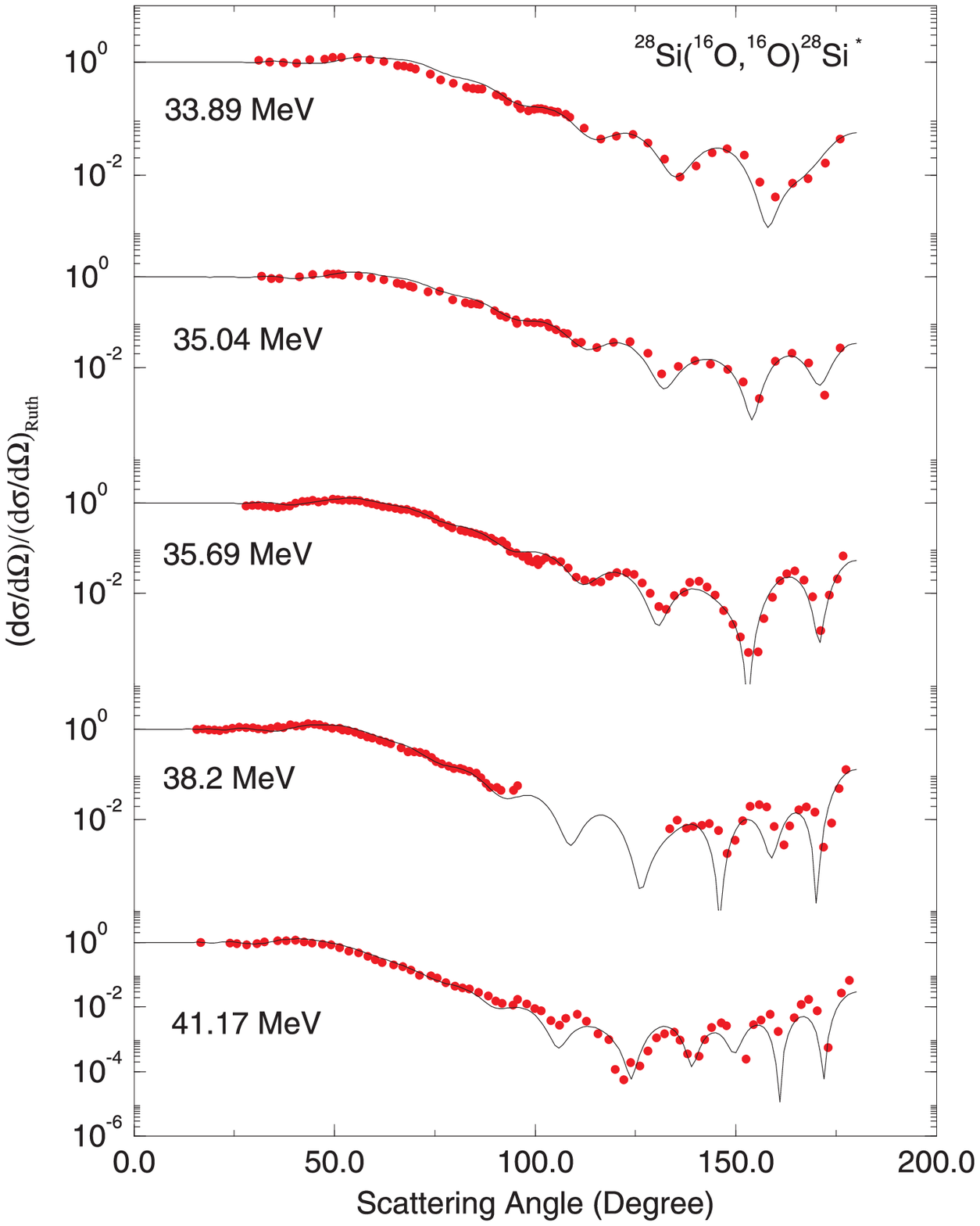}}  \vskip+0.5cm
\caption{Ground state results obtained using the new coupling
potential with the exact $\beta$ value ({\it continued from figure
\ref{2ndgroundosi}}).} \label{2ndground2osi}
\end{figure}

\begin{figure}[bt]
\epsfxsize 9.5cm \centerline{\epsfbox{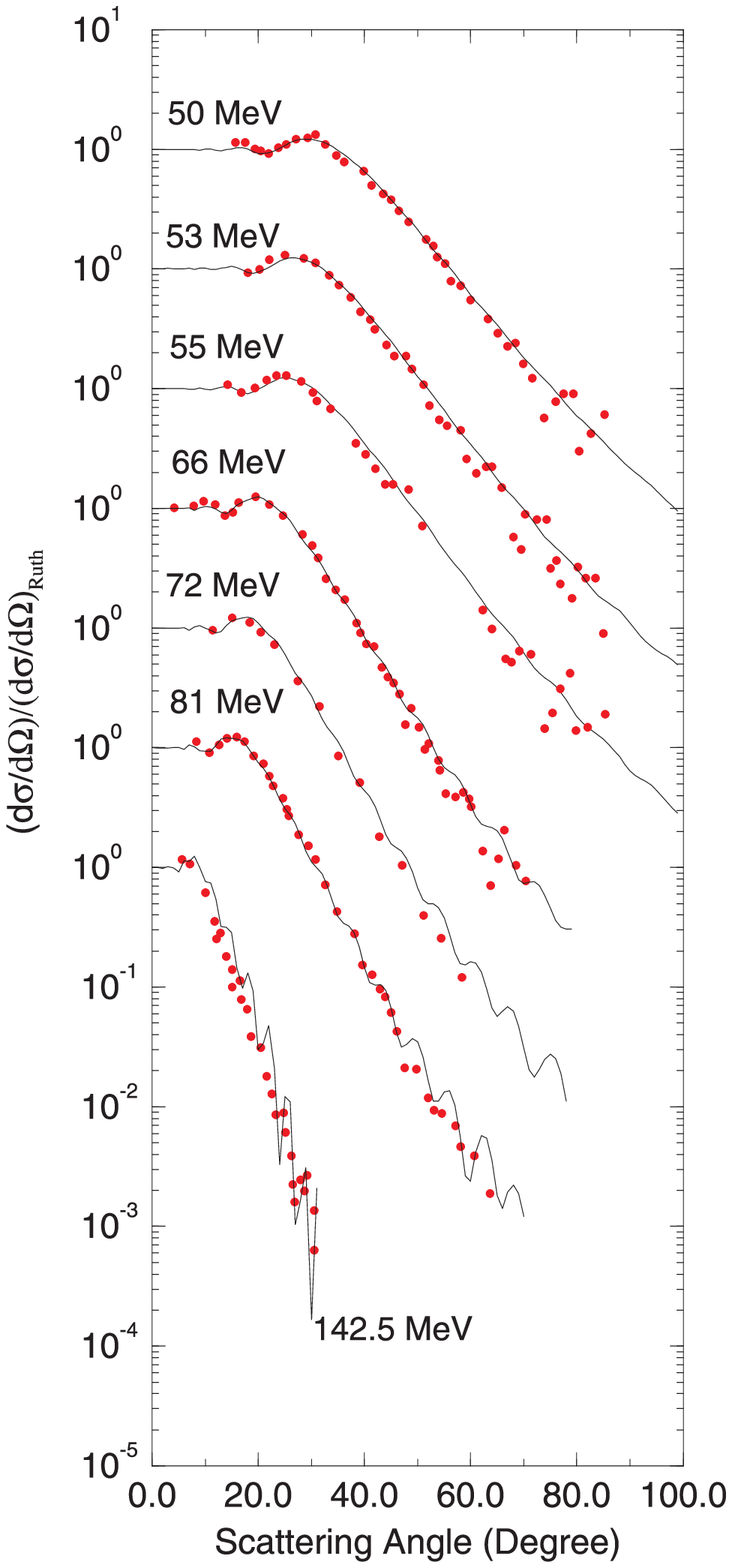}} \vskip+0.5cm
\caption{Ground state results obtained using the new coupling
potential with the exact $\beta$ value ({\it continued from figure
\ref{2ndground2osi}}).} \label{2ndground3osi}
\end{figure}

\begin{figure}[bt]
\epsfxsize 13.5cm \centerline{\epsfbox{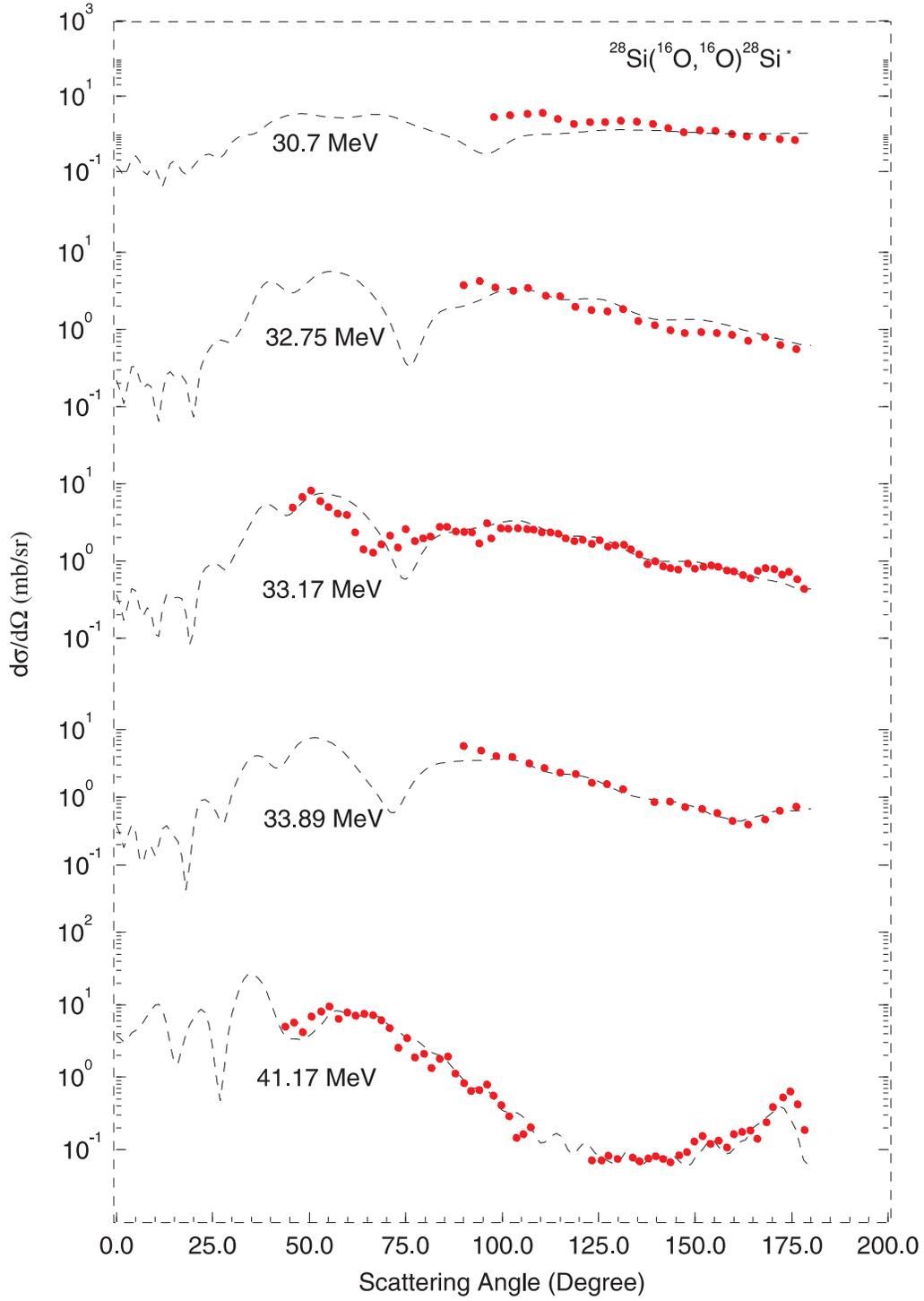}}  \vskip+0.5cm
\caption{2$^{+}$ excited state results obtained using the new
coupling potential with the exact $\beta$ value.} \label{2ndin}
\end{figure}

\begin{figure}[bt]
\epsfxsize 9.5cm \centerline{\epsfbox{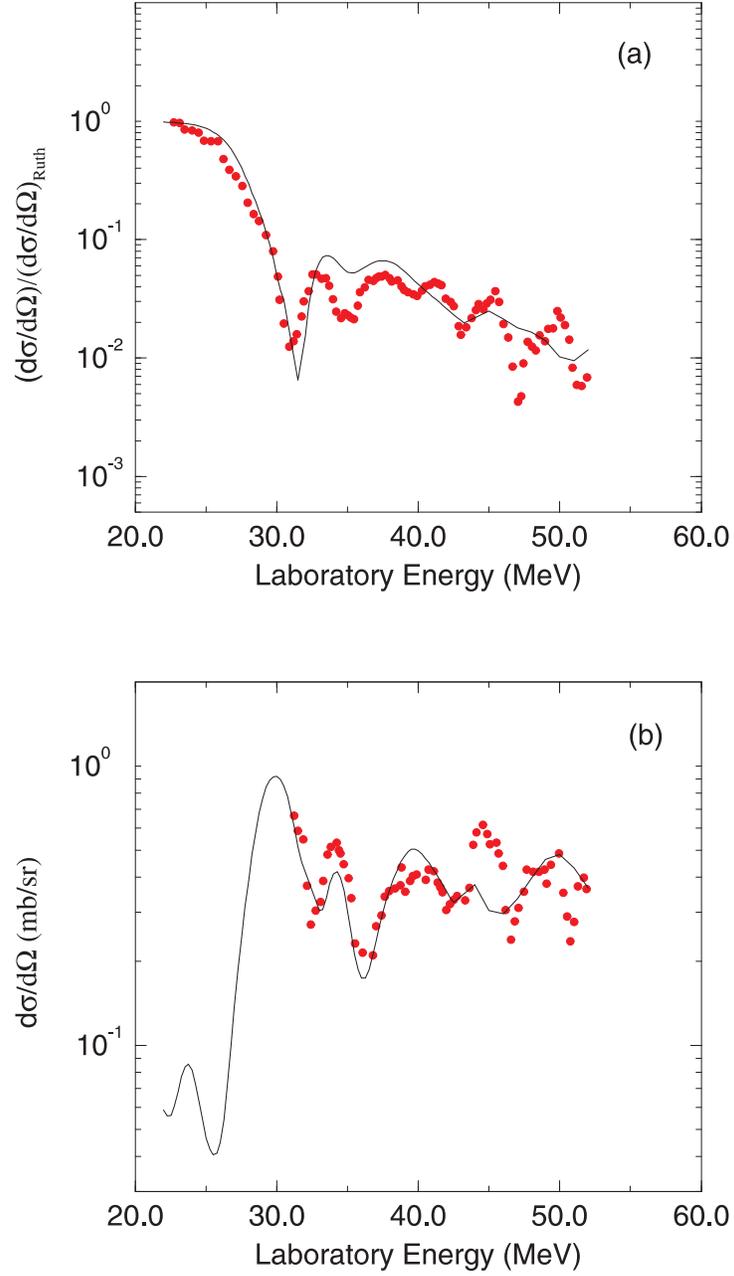}} \vskip+0.5cm
\caption{180$^{\circ}$ excitation function results obtained using
the new coupling potential with the exact $\beta$ value for $(a)$
the ground and (b) 2$^{+}$ states.} \label{2ndexc}
\end{figure}

We have analyzed the experimental in the same energy range using
this empirical $\beta_{2}$ value and the results of the new
coupled-channels calculations are shown in figures
\ref{2ndgroundosi}, \ref{2ndground2osi} and \ref{2ndground3osi}
for the ground state. Figure \ref{2ndin} presents the inelastic
scattering while figure \ref{2ndexc} shows the 180$^{\circ}$
excitation function for the ground and first excited states.

This new coupling potential solves the out-of-phase problem as
shown in figure \ref{2ndexc} and fits the ground state data and
the 180$^{\circ}$ excitation functions simultaneously. A
comparison is given for the 2$^{+}$ excited state in figure
\ref{2ndexc}. While the standard coupling potential is
out-of-phase with the measured one, this new coupling potential
significantly improves the agreement with the experimental data
and solves the out-of-phase problem.

It is striking that the phase variation and the absolute magnitude
of the inelastic cross-sections for all energies are correctly
accounted for with this model. In contrast to the predictions of
the standard coupled-channels calculations, the magnitude of the
2$^{+}$ excited state data at lower and intermediate energies
where we have available experimental data for the individual
angular distributions are fitted well. The comparison of the
$\chi^{2}$ values with the standard one is given in
table~\ref{chi}.

Finally, table \ref{volint} shows the volume integrals of the real
and surface and volume imaginary potentials. The volume integrals
of the real and imaginary potentials are calculated by using
following formulae:
\begin{eqnarray}
J_{V}(E)=\left[\frac{4 \pi}{A_{P}A_{T}}\int_{0}^{R}V(r,E)r^{2}dr
\right]
 \nonumber \\
 \nonumber \\
J_{W}(E)=\left[\frac{4 \pi}{A_{P}A_{T}}\int_{0}^{R}W(r,E)r^{2}dr
\right]
\end{eqnarray}
The radii of the imaginary potentials are calculated from
equations \ref{volr} and \ref{surr}. It is seen from table
\ref{volint} that the potentials fulfill the dispersion relations
and the agreement between r$_{W_{S,V}}$ and J$_{W_{S,V}}$ is very
good.
\section{Summary}
\label{conc}
We have shown a consistent description of the elastic and
inelastic scattering of the $^{16}$O+$^{28}$Si system from 29.0
MeV to 142.5 MeV in the laboratory system by using the standard
and new coupled-channels calculations. In the introduction, we
presented the problems that this reaction manifests. We attempted
to find a consistent solution to these problems. However, within
the standard coupled-channels method, we failed, as others did, to
describe certain aspects of the data, in particular, the magnitude
of the 2$^{+}$ excitation inelastic scattering data although the
optical model and coupled-channels models explain perfectly some
aspects of the elastic scattering data. In order to reproduce the
first excited state (2$^{+}$) data in the standard
coupled-channels calculations, we were compelled to increase the
value of nuclear deformation and such arbitrary uses of $\beta$
have been practiced in the past without giving any physical
justifications other than stating it is required to fit the
experimental data. Although we obtained a reasonable agreement
between the experimental data and theoretical calculations for the
ground and 2$^{+}$ state data, the standard coupled-channels
method have totaly failed in providing simultaneous fits to the
individual angular distributions and 180$^{\circ}$ excitation
functions and could not solve the out-of-phase problem between the
theory and experimental data for these states.

We have, however, obtained excellent agrement with the
experimental data over the whole energy range studied by using a
new coupling potential, which has been outstandingly successful in
explaining the experimental data for the the
$^{12}$C+$^{12}$C~\cite{Boz1} and $^{12}$C+$^{24}$Mg~\cite{Boz2}
systems over wide energy ranges. The comparison of the results
indicates that a global solution to the problems relating to the
scattering observables of this reaction over a wide energy range
has been provided by this new coupling potential. However, it is
not possible at present to provide a solid theoretical foundation
and further work in order to derive this term from a microscopic
viewpoint is still under-progress. Any insights that would lead to
progress in this direction will be greatly welcome in the future.
\section{Acknowledgments}
Authors wish to thank Doctors Y. Nedjadi, S. Ait-Tahar, B. Buck,
A. M. Merchant, Professor B. R. Fulton and Ay\c{s}e Odman for
valuable discussions and encouragements. I. Boztosun also would
like to thank the Turkish Council of Higher Education (Y\"{O}K)
and Erciyes University, Turkey, for their financial support.
\begin{table}
\begin{center}
\begin{tabular}{lll}
$E_{Lab}$ &  Standard CC &  New CC  \\     \hline
29.34 & 2.5 &   3.7 \\
29.92 & 2.1 &   3.5 \\
30.70 & 3.0 &   3.4 \\
31.63 & 2.7 &   4.1 \\
32.75 & 1.7 &   2.7 \\
33.17 & 3.3 &   3.0 \\
33.89 & 3.2 &   2.3 \\
35.04 & 13.7 &  3.3 \\
35.69 & 13.5 &  8.9 \\
38.20 & 46.0 &  11.7 \\
41.17 & 127.0 & 29.4 \\
\end{tabular}
\end{center}
\caption{The numerical values of $\chi^{2}$ for the standard and
new CC cases in the inelastic scattering calculations.}
\label{chi}
\end{table}
\begin{table}
\begin{center}
\begin{tabular}{lll}
$E_{Lab}$ (MeV) &  $J_{W_{S}}$ (MeV fm$^{3}$)&  $J_{W_{V}}$ (MeV
fm$^{3}$)
\\ \hline
29.34 & 3.23 & 0.25 \\
29.92 & 3.59 & 0.27 \\
30.70 & 4.12 & 0.31 \\
31.63 & 4.81 & 0.35 \\
32.75 & 5.75 & 0.40 \\
33.17 & 6.13 & 0.42 \\
33.89 & 6.82 & 0.45 \\
35.04 & 8.03 & 0.52 \\
35.69 & 8.78 & 0.56 \\
38.20 & 12.10 & 0.73 \\
41.17 & 17.02 & 0.97 \\
\end{tabular}
\end{center}
\caption{The volume integrals of the surface and volume imaginary
potentials for the new coupled-channels calculations. The volume
integral of the real potential is 381.9 MeV fm$^{3}$}
\label{volint}
\end{table}
\end{document}